 \documentclass[preprint]{aastex701}

\usepackage[english]{babel}
\usepackage{natbib}
\usepackage[letterpaper,top=3cm,bottom=2cm,left=3cm,right=3cm,marginparwidth=1.75cm]{geometry}

\usepackage{amsmath}
\usepackage{amssymb}
\usepackage{graphicx}
\usepackage{gensymb}
\usepackage{multirow}
\usepackage{wrapfig}
\usepackage{tabularx}
\usepackage{subcaption}
\usepackage{lineno}

\newcommand{\be}{\begin{equation}}
\newcommand{\ee}{\end{equation}}
\newcommand{\nn}{\mbox{} \nonumber \\ \mbox{} }
\newcommand{\ba}{\begin{eqnarray}}
\newcommand{\ea}{\end{eqnarray}}

\newcommand{\Alfven}{Alfv\'{e}n }

\newcommand{\curl}{{\rm curl\, }}
\newcommand{\A}{{\bf A}}
\newcommand{\E}{{\bf E}}
\newcommand{\B}{{\bf B}}
\newcommand{\J}{{\bf J}}
\renewcommand{\v}{{\bf v}}
\renewcommand{\k}{{\bf k}}

\newcommand\eg{{\it{e.g.\ }}}

\newcommand{\Bf}{{magnetic field}}
\newcommand{\Bfs}{{magnetic fields}}

\begin{document}

\title{Dynamics and stability of magnetized AGN-blown bubbles in clusters of galaxies}

\author[gname=Aleksey,sname=Mohov]{Aleksey Mohov}
\affiliation{Purdue University}
\email{amohov@purdue.edu}
\author[gname=Maxim,sname=Lyutikov]{Maxim Lyutikov}
\affiliation{Purdue University}
\email{lyutikov@purdue.edu}

\begin{abstract}

We perform MHD simulations of AGN-blown bubbles in  the Intercluster Medium (ICM) containing large-scale coherent magnetic fields.  
We assume that bubbles, created by the intermittent jets from Active Galactic Nuclei,  quickly relax to the Woltjer-Taylor  spheromak-like state, with internal plasma beta-parameter  $\sim 1$. We demonstrate that such  bubbles  rising through hydrostatically-stratified  atmosphere are magnetically   stabilized against fluid interface  instabilities, remaining coherent for a long time. Typical   velocity  is $ v /c_s \sim \sqrt{R/H} \leq 1 $ ($c_s$ is sound speed, $R$ is the bubble size, $H$ is the scale height). Current-driven instabilities (internal kinks) lead to bubble's tilting, but  develop on long time scales,  and  remain unimportant, leading to minor modifications of the internal structure.  Our results explain apparent long-term stability of ICM cavities. Subsonically rising stable bubbles dissipate in their wake approximately the energy initially injected by the jet, and may efficiently reheat the clusters cores in a ``gentle'' way.

\end{abstract}
\section{Introduction} One of the most striking features of galaxy cluster  is the presence of numerous, large radio cavities in the ICM plasma. The Chandra-X mission observed numerous such bubbles \citep{Jets}, \eg in the prototypical case of Perseus Galaxy Cluster \citep{ChandraX}.  Black hole jets are believed to carve out these bubbles and velocity measurements show them to be moving away from the central gravitational potential, i.e. floating (see e.g. \cite{10.1111/j.1365-2966.2008.13027.x}).


Typical dynamical time scale of jets carving the bubbles is $\sim 10-100 Myrs$.
This is much longer that the dynamic times scale of the inner part of the cluster of few million years (for scale height $H \sim $ few kpc and temperatures $\sim $ few $10^7$ K \citep{1984ApJ...276...38J}).
Thus, observations of numerous bubbles in a given cluster imply long survival times for a given bubble, much longer than the dynamical time. 
  
In contrast, prior {\it hydrodynamic} simulations show that the bubbles are  quickly disrupted by hydrodynamic instabilities on the scale-height crossing timescale \citep{Churazov,2015ApJ...815...41R}. 
Steady cosmic ray support \citep{2019ApJ...871....6Y}, and Braginskii viscosity \citep{2019ApJ...883L..23K} have also been explored as stabilizing mechanisms, but cannot keep the bubbles stable across instabilities of all length scales. The onset of instabilities can be delayed if the bubble is inflated by a slow, wide angle jet, creating a dense 'shell' around the bubble \citep{2006MNRAS.371.1835P,2008MNRAS.389L..13S}; nevertheless the question remains open.
This is the problem we address in the present work: hydrodynamic simulations show quick disruption of bubbles, while from their frequency and motion, we conclude that the bubbles must be long-lived (in terms of $H/c_s$, the pressure scale height divided by the sound speed), and able to rise in an atmosphere with reasonable coherence. Previous stability simulations of ICM cavities have been primarily hydrodynamic. \cite{Robinson} outline the problem: An purely hydrodynamic bubble will splinter into many smaller pieces on a short timescale. Their simulations are primarily 2-D, but show already the necessity of other stabilizing mechanisms. They included a simple magnetic field of azimuthally oriented flux tubes as an example of such a mechanism.

Lacking a stable MHD bubble, other methods have been used to explore the interactions between  the bubble and it's environment.\citep{2000A&A...356..788C} showed that   Sheared/vortical velocity structure of the purely hydrodynamic  bubbles may  contribute to stability. \cite{Congyao} simulated a rigid rising bubble in a hydrostatic atmosphere. While this prior study focused on mapping the eddies and wakes of the rising bubble, it notes specifically the limitations on a) the small scale 3-D turbulence and b) the deformation of the bubble itself as it rises.

With a fully fluid 3-D domain and no 'rigid' approximations, we can fulfill both of these roles implicitly. Our work improves on both prior paradigms with a novel and physically plausible bridge between the regime of a quickly destroyed bubble, and the studies conducted with an inserted rigid bubble.

\begin{figure}[h!]
 \includegraphics[width=.99\linewidth]{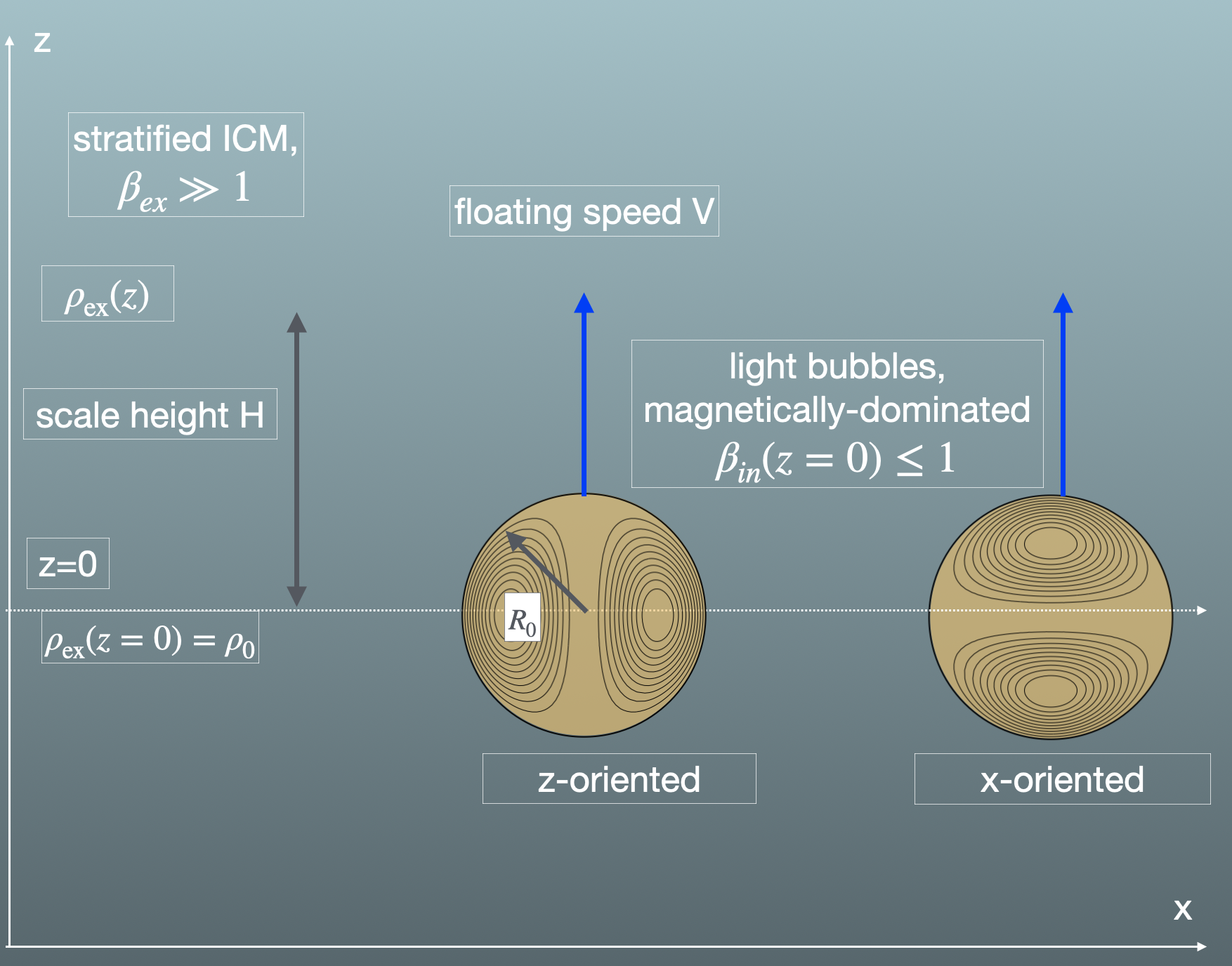}
\caption{Cartoon of the mode. Magnetized bubble, modeled as  pressure-confined, mildly  magnetized $\beta_{0,in} \leq 1$ spheromaks.  Initially, the bubbles has radius $R_0 \leq H$,  The bubble is submerged into  unmagnetized, gravitationally  stratified  ICM.   Two  cases of \Bf\ orientation within the bubble,  horizontal  and vertical,  are investigated.}
\label{fig:overview}
\end{figure}

The main new ingredient in the present work  is the  assumption of large-scale, coherent internal \Bfs,  Fig. \ref{fig:overview}, and Appendix \ref{pressure-confined}. \cite{2002ARA&A..40..319C} note the presence of magnetic fields throughout the bulk of the ICM, and that there is significant spatial variation. We assume that after termination of an active jet, a newly created, pressure-confined magnetic bubble relaxes following the  Woltjer-Taylor relaxation  principle \citep{Woltjer58,1974PhRvL..33.1139T}. 
Relaxation of complex, helical magnetic structures into force-free spheromak-like shape has been observed in the laboratory  \citep{PhysRevE.97.011202}. Nevertheless, the transition from black hole jet to cavity remains an open question, but is not the subject of our work here.

In the ICM, the relaxation will occur on  \Alfven time scale of $t_A \sim$ few Myrs, for $R\sim H$, \Bf\ $\sim \mu$Gauss and hydrogen gas density $n\sim 10^{-3}$ cc$\Rightarrow \rho\sim10^{-27}$ g. Explicitly, $v_a \sim 10^{-6}/\sqrt{10^{-27}}=10^{7.5}$ cm/s and $t_a \sim 10$ kpc/$v_a \sim 1$ Myr ({\Bf\ relation $t_A \sim t_d$ follows from the assumption of internal beta-parameter $\sim 1$ and bubble size $\sim H$}). 

As  an analytical approximation to the relaxed configuration, we use spheromak-like structure of pressure-confined magnetic bubble with  no-surface-current  \citep{2010MNRAS.409.1660G}, see Appendix  \ref{pressure-confined}. Total \Bf, both toroidal and poloidal componnets, are zero on the surface. The  solutions can be generalized to include  finite external  \Bfs, see Appendix \ref{pressure-confined}.

\section{Rising bubbles: theoretical expectations}


\subsection{Local dynamics of fluid bubbles}
\label{local}
Even if effects of the \Bf\ are neglected, The fluid dynamics of bubbles rising in the ICM occurs in a very different regime than more studied air bubbles in water  \citep[\eg][]{2015NatCo...6.6268T}. In the case of ICM bubbles, effects of surface tension and viscosity are completely negligible \citep[Reynolds, Weber, Galilei, Bond,  and E\"{o}tv\"{o}s numbers  are all much larger than unity][] {Lamb}. The fluid drag then occurs in  Reynolds, not Stokes, regime \citep{LLVI}. (While we expect the astrophysical system to have no viscosity, it is possible that numerical viscosity could affect our results, but we do not believe that has occurred in this work. A higher resolution convergence study could confirm this.)
Another difference from air bubbles rising in water is the fact that water is nearly incompressible, of constant density - this corresponds to very large sound speed. The closest hydrodynamics analogy is the dynamics of large air bubbles \citep{1950RSPSA.200..375D}.

For bubbles of not very large size, $R  \leq H$, with rising speeds $V \ll c_s $ (to be confirmed), a bubble is approximately in pressure balance; there is a  slight top-bottom  disbalance that provides a buoyant force. In this work we neglect the effects of external \Bf, see \S \ref{hovering} and Eq. (\ref{beta1}), explicitly solving for 0 \Bf\ on the surface of the bubble and in the ambient atmosphere.

In this approximation, the motion of a bubble of a given size $R$  and internal density $\rho_{\rm in}$  rising through stratified external medium   of density $ \rho_{\rm ex}$ is determined by buoyant force $F_b$
\be
F_b \approx   g    (\rho_{\rm ex} -\rho_{\rm in} )  R^3
\ee
and  
fluid drag $F_{drag}$ (in the limit of large Reynolds numbers)
\be
F_{drag} \approx c_1\rho_{\rm ex} R^2 V^2
\label{Fdrag}
\ee
($g$ is a local acceleration due to gravity, $c_1$ is a coefficient of the order of unity). 

Buoyant and drag forces determine acceleration;  importantly 
 for a  bubble with given radius $R$ the relevant  dynamical quantity is   the  ``added'' mass \citep{LLVI}
\be
\delta M = \frac{2 \pi}{3} R^3 \rho_{\rm ex}  \approx \rho_{\rm ex}  R^3
\ee
Added mass contributes to force balance {\it only} if a bubble accelerates. 

As a result, in the limit $\rho_{\rm in} \ll \rho_{\rm ex}$, very quickly the bubble reaches constant velocity
\ba && 
V \approx  \sqrt{ g R}  = c_s  \sqrt{ R/H }
\nn &&
H = c_s^2/g
\label{hydro-expect}
\ea
where $H$ is the scale height \citep{1950RSPSA.200..375D}.
This corresponds to Froude number $Fr\approx 1$.

Thus,  we expect that the bubble's velocity increasing with bubble's size as   $\sqrt{ R}$. Bubbles as large as scale height are expected to rise with nearly sonic velocity. We find the magnetized bubbles conform to this scaling relation well independently of initial radii (See Fig. \ref{fig:rising_speeds}).

\subsection{Disruption of fluid bubbles}

It is expected that fluid bubbles are disrupted by KH instability on scales
\ba &&
\tau_d \sim R/V = \frac{ \sqrt{R H}}{c_S} =  \sqrt{ \frac{R}{g}} 
\nn && 
\Delta z = V \tau_d = R \ll H
\ea
Hence, very fast, after moving approximately the distance equal to the size. This simple estimate agrees with numerical calculations and work in e.g. \cite{Robinson}.

Due to the density difference inside and outside the bubble, we also expect (and observe in purely fluid simulations) the development of the Rayleigh-Taylor Instability, with the characteristic growth rate $\sigma$:

\begin{equation}
    \sigma^2 = \frac{\rho_{in}-\rho_{out}}{\rho_{in}+\rho_{out}}
\end{equation}

 The introduction of a magnetic field suppresses this instability and creates a critical wavelength: $\lambda\sim B^2/g(\rho_{in}-\rho_{out})$, as in, e.g. \citep{2019arXiv190511650K}.
\subsection{Magnetized bubble in stratified medium}
\label{hovering}

Above, in \S \ref{local}, we discussed local velocity of a  fluid  bubble of given size $R$ in stratified external medium. Next we discuss temporal evolution of bubble's properties, including effects of large-scale internal \Bf.

The evolution of the parameters of rising bubble is determined by the conservation laws (mass and, most importantly, magnetic flux), equation of state of the internal material, and external density and pressure  structure. Internal and external media may have different EoS, and different magnetizations. The general set-up is very parameter-rich.

As an important simplification, 
we assume that the bubble is mildly magnetized, and  
neglect effects of the \Bf\ in the external medium. 
The corresponding plasma beta-parameters are
 \ba &&
 \beta_{\rm ex} = \frac{ p_{\rm ex} }{B_{\rm ex}^2 /( 8 \pi)} \gg 1
 \nn &&
 \beta_{\rm in} = \frac{ p_{\rm in}}{B_{\rm in}^2 /( 8 \pi)} \lesssim 1
 \label{beta1}
 \ea

 The two media, internal and external,  may have different EoS, parametrized by adiabatic indices $\gamma_{\rm in}$ and  $\gamma_{\rm ex}$. For example, if the bubble is filled with relativistic particles, $\gamma_{\rm in} =4/3$, while for the non-relativistic external gas  $\gamma_{\rm ex}=5/3$.

 As a bubble rises, its size, internal \Bf\ and pressure, as well and internal  beta-parameter $ \beta_{\rm in} $ change. 
For bubbles rising slowly (with velocity much smaller than the sound speed),  the external-internal pressure balance reads
 \be
 p_{\rm ex}\approx  B_{\rm in}^2/(8\pi)  + p_{\rm in} 
 \label{pressure}
\ee
This condition is satisfied approximately, as the difference between external pressure at the base and the top of the bubble gives rise to buoyant force. In addition, for large scale internal \Bf, internal hoop stresses contribute locally to the force balance.

To address the basic properties of rising bubbles, in this paper we consider a plane-parallel atmosphere with isothermal equation of state $\gamma_{\rm ex}=1$.  The external  pressure  then scales as
\ba &&
p_{\rm ex} = p_0 e^{- z/H}
\nn &&
p_0 = c_s^2 \rho_0
\nn &&
H=\frac{c_{s}^2}{g}
\ea
where $g$ is an effective gravity, $\rho_0$ is external density at $z=0$ and $c_s$ is a global constant. The hight $z$ is measured from the initial bubble injection point.

In addition, for numerical simplicity, we model the internal plasma as isothermal, with the same EoS as external medium. This paradigm leaves one degree of freedom as the magnetic forces and pressure gradient forces must balance inside the bubble. An extension to varying EoS inside and outside (and use of an adiabatic EoS) the bubble would allow for an extra degree of freedom i.e. 'hot' and 'cold' bubbles. Given the exploratory nature of this work, such complexities were deemed unnecessary at this time. It is the presence of the internal \Bf, which partially compensates the external pressure, and makes the bubble lighter, Eq. \ref{pressure}. We expect that more realistic EoS will result in primarily quantitative differences from the isothermal case, with our results extending generally to more general scenarios.

Conservation of mass  and magnetic flux imply
\ba && 
R^3  \rho_{\rm in}=  \rho_{0, \rm in} {R_0^3}
\nn &&
 B  _{\rm  in}  R^2= B_0 R^2_0 
\label{mass}
 \ea
 (where subscripts $0$ indicate the initial values).

 The force-balance  (\ref{pressure}) then determines $R(z)$: 
  \be
   \left( \frac{R_0}{R} \right)^4 +  \beta_{0, \rm in} \left(\frac{R_0}{R} \right)^3 = (1+ \beta_{0, \rm in}) \exp\{  - z/H\},
   \ee


   As the bubble expands, its internal beta-parameter increases
   \be
    \beta_{\rm in} =  \beta_{0, \rm in} \left(\frac{R}{R_0} \right) 
    \ee
    while the total  buoyancy-minus-gravity  force evolves according to 
   \be
   F_b= g R_0^3  \rho_0 \frac{1}{1+ \beta_{0, \rm in}} \frac{R_0}{R} \approx F_{b,0}  \frac{R_0}{R} 
   \ee
  where $F_{b,0}$ is the initial buoyancy force at $z=0$. 
 
 Though formally $F_b$ never becomes zero, the buoyancy force quickly reduces as the magnetized isothermal bubble expands in isothermal atmosphere. Qualitatively, when $ \beta_{\rm in}  \sim 1$ the mass of the bubble is of the order of the mass of external medium within the bubble's volume. The buoyancy force becomes negligible then.

 We find that the case of a perfectly magnetic bubble is not confirmed by simulations (for the given mild magnetization). The bubble rises slowly (much less than $c_S$ for $R \leq H$), but does expand and continue to rise. Thus the numerics show a balance between the purely hydrodynamic cases and the purely magnetically dominated cases. In a purely hydrodynamic case, we would expect fast expansion and rising through the medium, coupled with surface instabilities that destroy the bubble over a few dynamical times. In a purely magnetic case, a well confined bubble maintains its size and is expected to  hover after reaching some equilibrium point. Instead, we see a balance between the two that is qualitatively parameterized by the relative size of the bubble versus the strength of the "scaffolding" magnetic field. In the cases of the a smaller bubble with similar plasma $\beta$, the bubble will expand much less, and also rise more slowly. Conversely, larger bubbles exhibit expansion and faster rising.
 
 Notably, surface instabilities remain suppressed in all cases.  The stablizing effect of the magnetic field persists through a wide range of regimes, and is not dependent on the magnetic field being so strong as to completely dominate.

\section{Methodology}
\label{Methodology}

We seek to demonstrate that a relaxed spheromak-like magnetic field can serve as a framework to extend the lifetime of a plasma bubble. Beyond simply surviving, observations point to bubbles floating away from the central black hole of the galaxy cluster. Therefore, we simulate a bubble capable of rising in a stratified, plane-parallel atmosphere with constant gravitational potential. The bubble should float upwards, and maintain its shape for a reasonable amount of time. Establishing this basic case, we then explore the parameter space of size and orientation of the bubble. We characterize some important bulk dynamical properties (e.g. rising speed) and explore any errata arising from the embedded magnetic field.

We will use "bubble" to mean the localized underdensity in the ICM gas.  Critically, the bubble is lower than the local hydrostatic density, has a well-defined boundary (as opposed to a shallow gradient), and remains contiguous for multiple dynamical timescales. For the purposes of plotting the bubble's movement over time, we define $\Delta \rho = (\rho-\rho_{HSE})/\rho$ where $\rho_{HSE}$ is the local hydrostatic density as calculated from a simple exponential in the case of an isothermal atmosphere. Another definition we find useful for bulk bubble dynamics is defining a threshold plasma $\beta<10$ and taking any cells satisfying this condition to be "inside the bubble". Mean statistics of these cells are used to calculate location, size and velocity of the contiguous bubble.

In this work, we use the Athena++ MHD code, a Godunov-type method useful for general astrophysical fluid problems \citep{Athena}. We modify the intitial conditions of a cartesian box with constant gravitational potential to include a magnetic field and density profile of the desired shape. This field matches a spheromak configuration, with the addition of an extra toroidal component to force zero current on the surface. The presence of the magnetic field adds constraints to the fluid setup: flux conservation and the addition of Lorentz forces. Including the fields and equation of state, we have a closed, numerically solvable system to describe the dynamics of the magnetized plasma.

 We use a simple plane parallel domain with constant gravity and isothermal equation of state, as this captures the main dynamical features of the system. The equilibrium density profile is then: $\rho(z)=\rho_0e^{-gz/c_s^2}$. The isothermal sound speed $c_s=1$. For $\rho_0$, we use the $z$ dimension, such that the mean fluid density across the whole domain is approximately $1$. Simulations with no bubble included were run to ensure this fluid background is stable for all relevant timescales.
 
Our domain is initiated with outflow boundary conditions on the sides and reflecting boundary conditions at the top and bottom (relative to gravity). In order to prevent reflected waves from interfering with bubble dynamics, simulations regarding ICM bubbles typically use outflow boundary conditions at every face, e.g. \citep{2010MNRAS.409.1660G},\citep{Congyao}. However, in this work, we find the dynamical timescales of interest are longer than it takes for the rarefaction wave to destroy the hydrostatic equilibrium. Possible remedies to the rarefaction wave include custom boundary conditions, or simply making the computational domain tall enough, such that the effects of the rarefaction wave are delayed past any time scales of interest. For this work, we reduce the amplitude of the sound waves as much as possible (through initial force balance),and focus on maintaining the hydrostatic atmosphere indefinitely. 
There are nonphysical effects from the boundaries when the bubble reaches the top of the domain, but this was only observed in the fastest rising cases. We keep the outflow conditions on the side, noting that the overall mass loss from them is small.
\begin{table*}
    \centering
    \begin{tabular}{|c|c|c|}
        \hline
        \multicolumn{3}{|c|}{Code Inputs}\\
        \hline
        Parameter& Value & Description\\ 
        \hline
        $c_s$&1& Isothermal sound speed\\
        $g$&1& Constant gravity\\
        $X \times Y\times Z$& $4H\times 4H\times 16H$& Domain dimensions \\
        $[x0,y0,z0]$&$[0,0,1]$&Initial bubble center\\
        $\beta_{\rm in}$& $0.27$, Fiducially& Minimum plasma beta\\
        $\alpha$ & Varies & $\alpha$ in spheromak equations\\
        \hline
        \multicolumn{3}{|c|}{Derived quantities}\\
        \hline
        Parameter& Value & Description\\
        \hline
        $R$&$5.763/\alpha$&Radius for 0 surface current\\
        $\rho_0$&$Z$&Amplitude of hydrostatic density curve\\
        $\rho_{top}$&$\rho_0*exp(-g*(z0+R))$&Hydrostatic density at top of bubble\\
        \hline
    \end{tabular}
    \caption{Spatially independent input parameters}
    \label{tab:input_parameters}
\end{table*}

Given an isothermal equation of state, the only sources of pressure in our domain are magnetic and density-gradient. Therefore, in order to create an initially approximately stable system, we set up our magnetic fields to create net force balance with the gas atmosphere at the top of the bubble. As detailed earlier, we use a twisted spheromak magnetic field, such that the net surface field is 0. This allows for both 0 magnetic field outside and no surface currents, which could be computationally unstable. The field is finite everywhere, and azimuthally symmetric. It differs from a simple spheromak by the addition of an extra toroidal term. The seeding of the bubble in a hydrostatic equilibrium creates a pressure imbalance at the bottom. This imbalance buoyantly initiates the rising of the bubble. 

The $[0,0,0]$ point presents an impasse: While the magnetic field is analytic, C++ does not handle the $r \rightarrow 0$ limit well. We can patch the point with the analytic solution, but this solution is not robust to changes in bubble orientation, magnetic field strength, and other parameters. Therefore, we patch the spatial coordinate, rather than the magnetic field: $[0,0,0] \rightarrow [0,0,10^{-15}]$. This allows us to manipulate the magnetic field equation as needed without re-patching, while still preserving the axial symmetry of the spheromak. We do not believe this patch has a significant impact on the results.

In order to encode the spatially dependent magnetic field and density efficiently, we first separate out the spatial invariants, which are typically constants dependent on various input parameters. These include inputs: $c_s$, $g$, $\alpha$, $[x0,y0,z0]$ (the initial center of the bubble), the orientation of the spheromak,$\beta_{\rm in}$, and the overall dimensions of the computational domain. Because the magnetic field varies between some maximal value (approximately halfway between the center and surface of the bubble), we define $\beta_{\rm in}$ as the minimum plasma beta in the initial setup, with the real plasma beta varying spatially $\beta_{\rm in}<\beta_{local}<\infty$. For the runs in this work, $\beta_{\rm in} ~ 1$ initially, with variations in $\beta_{\rm in}$ causing relatively small changes in the dynamics of the system. 

We derive some quantities once per time step rather than once per cell per time step. $R$ is the initial radius of the bubble. $\rho_{top}$ is the hydrostatic density at the top of the initial bubble location: $\rho_{top} = \rho_0 e^{-g(z0+R)}$. For the zero surface current setup, $R=5.763/\alpha $. $j(\alpha R)$ is the Bessel function evaluated at $\alpha R$; also we define its derivative $j'(\alpha R)$. $c1$ is the scaling coefficient for the B field, and scales as $c1\propto \sqrt{\beta_{\rm in}}*\rho_{top}$. The proportionality constant is set so that the chosen $\beta_{\rm in}$ matches the value calculated in the code. The key outcome of the setup is that we achieve force balance within the bubble, with a force imbalance only on the top and bottom of the bubble necessary to propel it upwards. We plotted the initial forces present in the bubble to ensure local force balance. Computationally, our domain has a resolution of 128 cells/$H$, and integrators of the second order are used for time and spatial reconstruction. We let the simulations run for $15 H/c_s$, noting that beyond this, we would need to increase the computational domain to avoid edge effects. 

\begin{figure}
    \centering
    \includegraphics[width=0.45\linewidth]{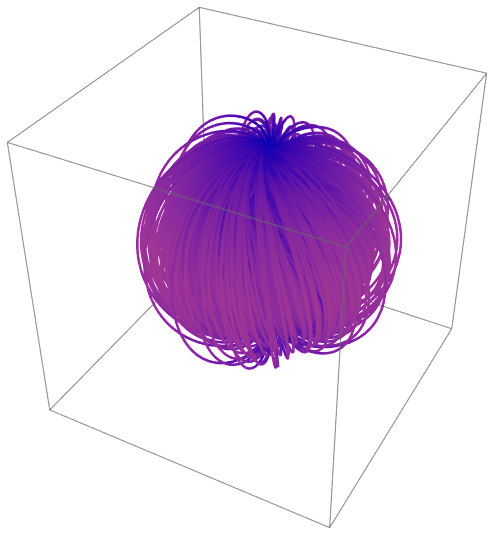}
    \includegraphics[width=.54\linewidth]{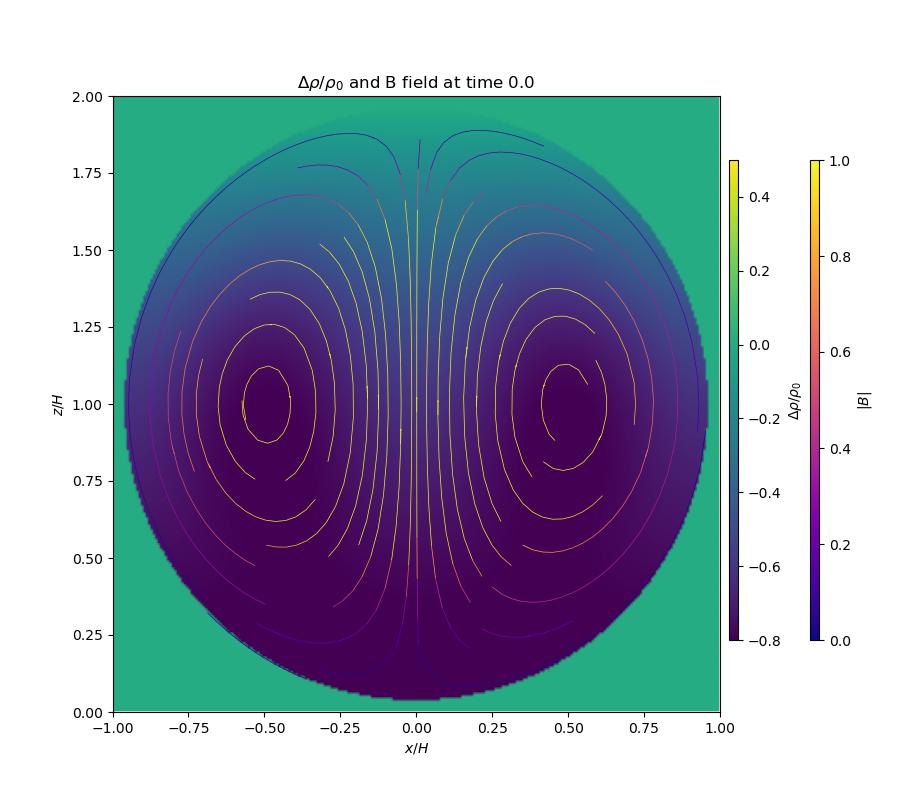}
    \caption{Pressure-confined magnetic bubble. Left panel:
    3-D rendering of the twisted  \Bfs\ embedded in the bubble,  resembling a "ball of yarn".
    Right panel: 2-D slice of initial  density variation $\Delta \rho/\rho$ and magnetic field streamlines. Note that the B field is axisymmetric, and any asymmetry in the streamlines are merely a plotting artifact. In the set-up the pressure is matched at the top of the bubble and is mildly  discontinuous elsewhere on the surface. Due to the external pressure gradient the bubble will experience a buoyant force propelling it upwards.}
    \label{fig:initial}
\end{figure}


We explore a number of permutations of our bubble to discern the key characteristics of its behavior. The initial $\beta_{\rm in}$ is kept constant at $\sim1$. We vary the $\alpha$ parameter, which in turn varies the initial radius ($R_0$)of the bubble. Varying the size of the bubble appears to have the most impact on rising speed and subsequent bubble dynamics. To cover interesting errata, we also have a run with the bubble initially horizontal. To understand how the bubble evolves in the absence of gravity, we include a few runs with no gravity and a cube shaped domain with outflow boundaries in every direction. This is also intended to diagnose any bias imparted by the reflecting boundaries necessary to maintain a hydrostatic atmosphere, in the presence of gravity. Lastly, we test a decoupling of $\alpha$ and $R$, which leads to a 'matryoshka' shape of nested bubbles. 

\begin{table}
    \centering
    \begin{tabular}{|c|c|c|c|c|}
        \hline
        Run & $\alpha$ & $R(H)$ & $g$ & orient\\
        \hline
        $\alpha_6$& $6$& $0.96$  & $1$ & $z$ \\
        $\alpha_{17}$ &$17.23$&$0.33$ & $1$ &$z$\\
        $\alpha_{25}$ &$25$&$0.23$ & $1$ &$z$\\
        $\alpha_{17x}$ &$17.23$&$0.33$ & $1$ &$x$\\
        $\alpha_{6g0}$ &$6$&$0.96$ & $0$ &$z$\\ 
        $\alpha_{17n}$ &$17$&$0.75$ & $0$ &$z$\\
        \hline
    \end{tabular}
    \caption{Parameter Space of Runs}
    \label{tab:parameters}
\end{table}

Representation of the results generally falls into two categories: visual representations of the entire domain and mean statistics about the bubble. 

For the former, we are interested in tracking the evolution of the fluid density and magnetic field over time. We can track the density in each cell and magnetic field streamlines, and this is used in the $g=0$ cases. For the runs with gravity, we instead plot magnetic field lines and the percentage deviation from local hydrostatic equilibrium density: $\Delta \rho = (\rho-\rho_{HSE})/\rho$, where $\rho_{HSE}$ is simply the exponential decay evaluated for local $z$. Typical values for $\Delta \rho$ range from $0.2$ to $-0.8$, i.e. a $20\%$ overdensity to an $80\%$ deficit. In the cases with gravity, the sound waves coming from the initial equilibration phase, while small in amplitude, can be visually distracting. Therefore, we also use $\log(\beta)$ to better distinguish the bubble in some cases. It is important to note that the spatial domain goes up to $z=16 H$, to help mitigate the effects of the upper reflecting boundary, however most interesting dynamics are contained in the presented $0H<z<8H$.

\section{Results} 

\subsection{Self-tilting}

Running prototypes of the bubble code, without gravity/external pressure  gradient, we encountered a slowly evolving, non-disruptive internal kink  instability (see Appendix \ref{Suydam}). This asymmetry causes the bubble to self tilt in a random direction.  Eventually, the magnetic field reconnects across the central axis. This reconnection then propagates outwards, destroying the spheromak and later the bubble as a whole. The entire destruction takes $\sim30H/c_s$, i.e. very slowly. The large-scale implication is a tilting of the whole bubble structure, both the field and the coupled fluid underdensity. We explored the tilting in detail to ensure the effect was physical rather than numerical.

\begin{figure*}[h!]
    \centering
    \begin{subfigure}{.32\textwidth}
        \includegraphics[height=.9\linewidth, width = \linewidth]{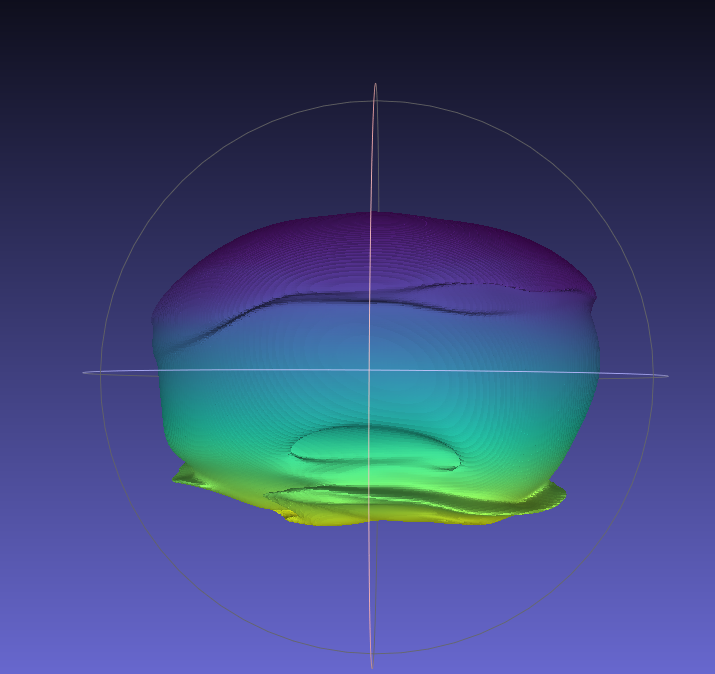}
    \end{subfigure}
    \begin{subfigure}{.32\textwidth}
        \includegraphics[height=.9\linewidth, width = \linewidth]{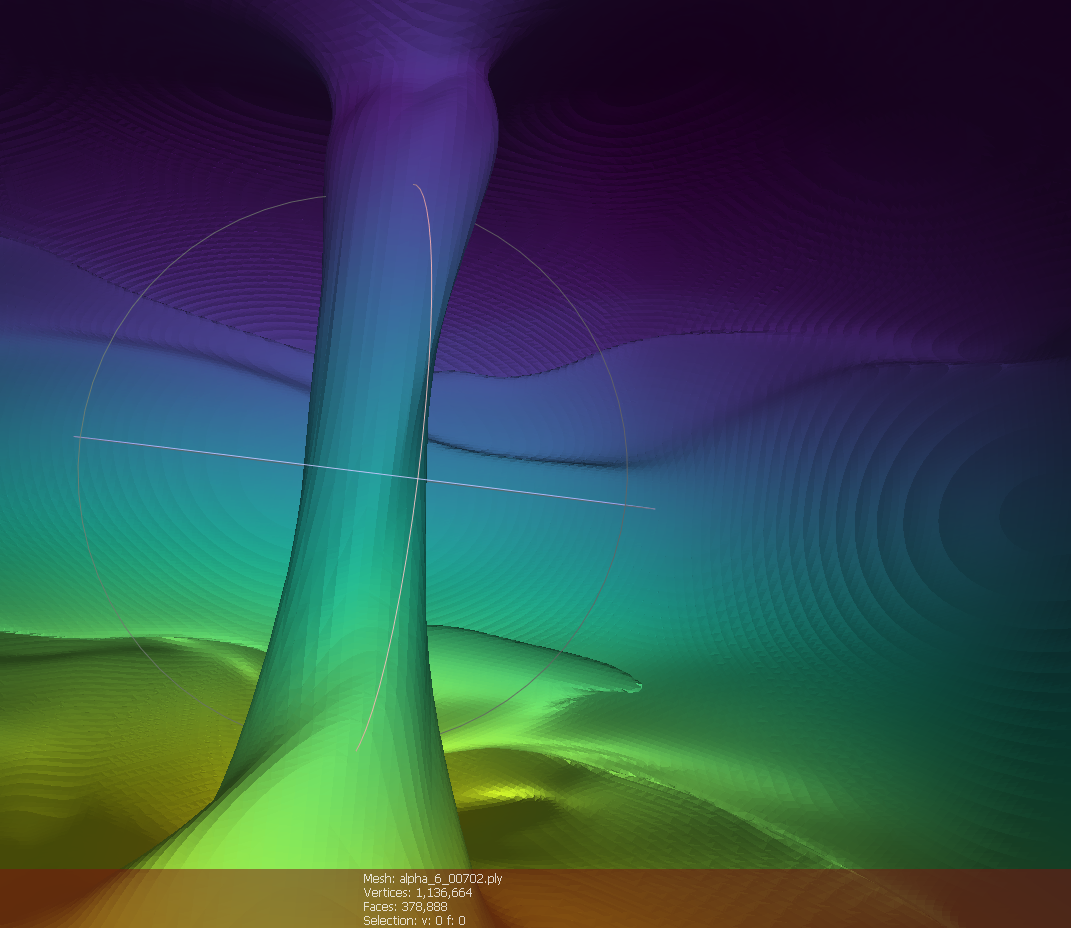}
    \end{subfigure}
    \begin{subfigure}{.32\textwidth}
        \includegraphics[height=.9\linewidth, width = \linewidth]{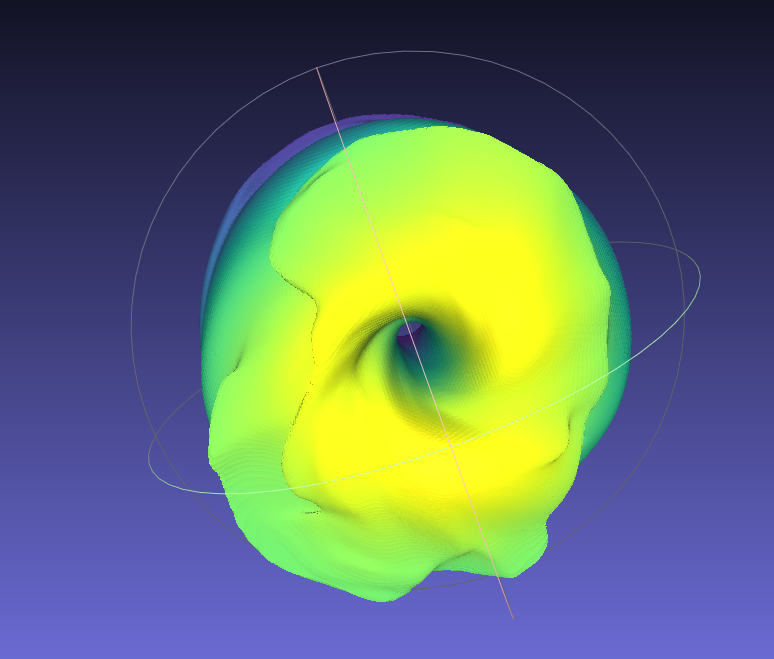}
    \end{subfigure}
    \caption{Renderings of $\beta=10$ isosurfaces of a buble of initial radius $R=.96H$. Left to right: view from the  outside,  zooming in from the inside to the bubble's core (note the kink-twisted structure),  and viewing  bottom-up. From the internal view, we can see the tight twisting and the  resulting asymmetry due to kink instability. The bubble also acquiers sheared toroidal velocity \citep{2011SoPh..270..537L}. Here color is just the z coordinate.}
    \label{fig:meshlab}
\end{figure*}


Understanding the tilting of the bubble under a variety of conditions requires recreating the effect under different of starting conditions. Tilting was observed in a number of cases: varying initial radius, varying initial plasma $\beta$, varying orientation. The key dichotomy is whether the cause of this tilting is a physical effect of the magnetic field setup, or a numerical artifact. In addition to the parameter space explored elsewhere, we also shifted the location of the entire bubble by less than a cell length, such that the $[0,0,0]$ point is removed from the spheromak equation, and reversed the azimuthal field direction. Both resulted in the same tilting instability. 

Lacking any definitive numerical culprit, we can loosely parameterize the tilting by its timescale. In the event of no magnetic field, one would expect no tilting, and indeed we find that the characteristic timescale to start tilting scales with initial $|B|$. Additionally, the tilting takes longer to manifest if $R$ is larger, due to crossing time needed to propagate outwards from a central instability. Combining, $\tau \propto (|B|/R)^\kappa$, where $\tau$ is timescale and $\kappa$ is some scaling exponent. 

The tilting is not fast on the scale of the rising bubble in atmosphere. Gravity may have a stabilizing effect to modulate the tilting as the bubble rises. Nevertheless the tilting remains important in the very long scale dynamics. 

In a run with no gravity, and outflow boundary conditions, the spheromak-like large scale magnetic field survives the tilting for $\gtrsim{20}$ dynamical times. Notably, the localized density deficit of the fluid remains indefinitely. Were the internal kink instability dampened entirely, we believe the spheromak field would remain stable indefinitely as well.

To further understand the dynamics of the central axis, we employed Purdue's Envision Center to observe the plasma $\beta$ isosurfaces in a 3D virtual-augmented reality environment (Fig. \ref{fig:meshlab}). In the 3D space, we observed the kink instability forming at the central axis of the bubble. Specifically, the spiraling magnetic field near the central axis of the spheromak twists tightly until it forms a kink, removing axial symmetry of the bubble and slowly propagating an instability outward. Importantly, this instability is both slow, and distinct from the primarily surface originating instabilities known to destroy ICM plasma bubbles. With the tilting erratum thus investigated, we moved on to the large scale dynamics of the bubble in atmosphere.

\subsection{Bubbles in stratified   atmosphere}

The first and foremost observation  is that  bubbles carrying coherent large-scale  \Bfs\   and rising through the stratified atmosphere  remain contiguous, for a range of   initial bubble radii, orientations and magnetization. The modified spheromak is capable of both supporting the bubble against the pressure gradient at its edge as well as suppressing any instabilities, e.g. Kelvin-Helmholtz, Rayleigh-Taylor, on the surface. 

The minimum  internal plasma beta, averaged over the whole structure is $\sim 1$. This is limited by our universal EoS (same inside and outside), and the nature of the starting analytical solution (kinetic pressure may become zero for too small overall beta).
The relevant quantity for dynamics is loosely quantified as a ratio between the radius of the bubble and the magnetic field strength. The latter, has a disproportionate impact on simulation cost when the Alfvenic velocity is $v_{A}\gg c_s$. Therefore, the variation of bubble radius as our primary parameter captures varied dynamics effectively and efficiently. 

In the case of the B field being the dominant driver of dynamics, the bubble remains a constant size and tilts on a slow time scale with no surface instabilities. A purely fluid bubble expands as it rises in a hydrostatic atmosphere, to compensate the lower ambient pressure. The balance between these two effects is diagnosable by the expansion of the bubble as it rises, and both regimes have some impact. 

\begin{figure*}[h!]
    \centering
    \begin{subfigure}{.49\textwidth}
        \includegraphics[width=\linewidth]{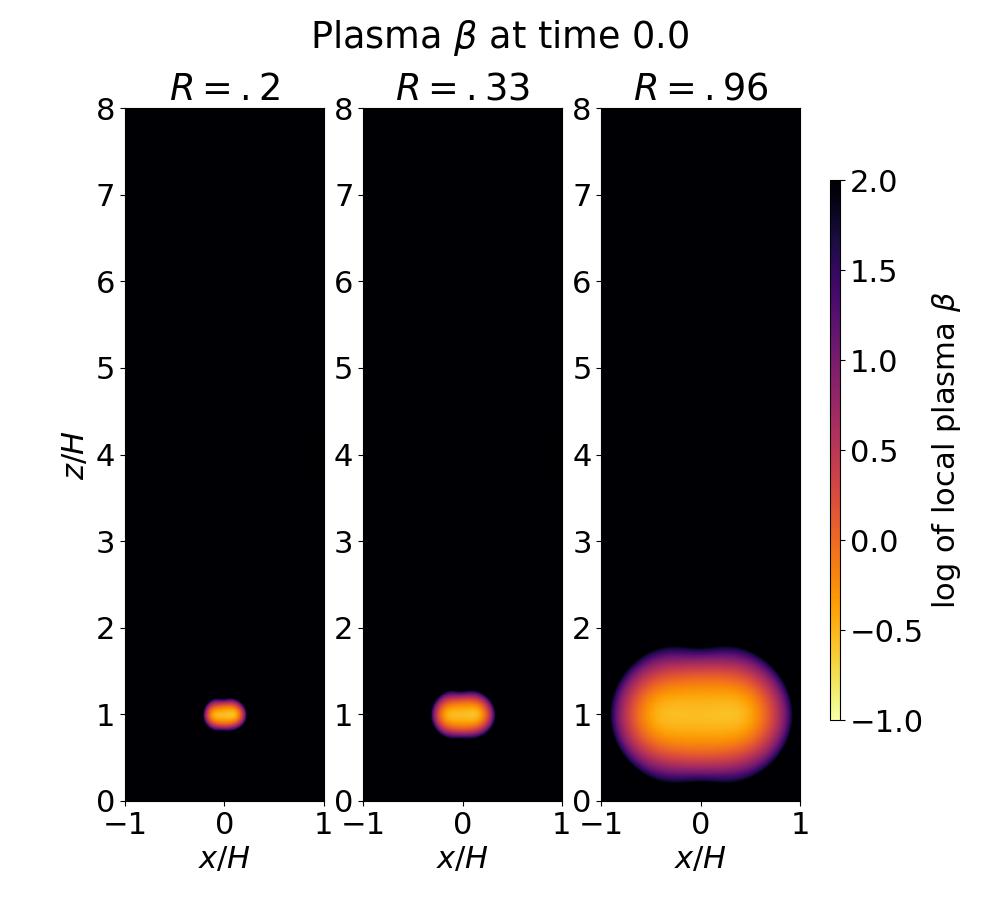}
    \end{subfigure}
    \begin{subfigure}{.49\textwidth}
        \includegraphics[width=\linewidth]{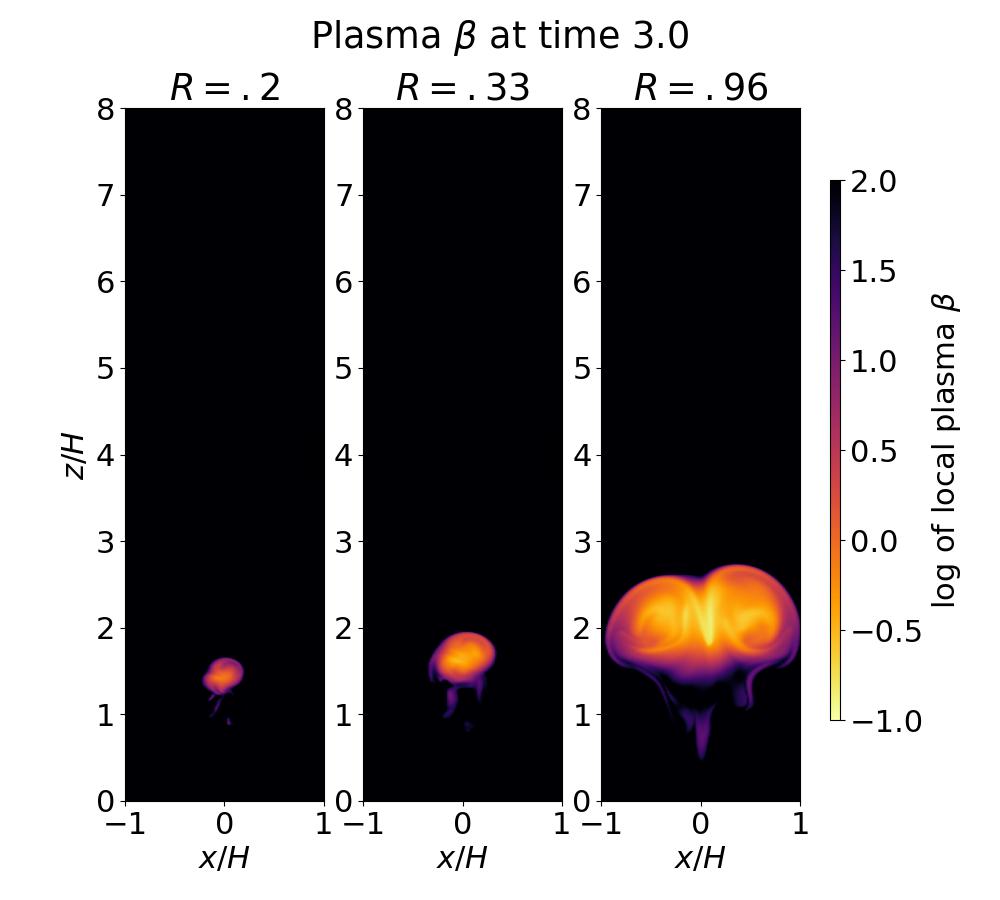}
    \end{subfigure}
    \begin{subfigure}{.49\textwidth}
        \includegraphics[width=\linewidth]{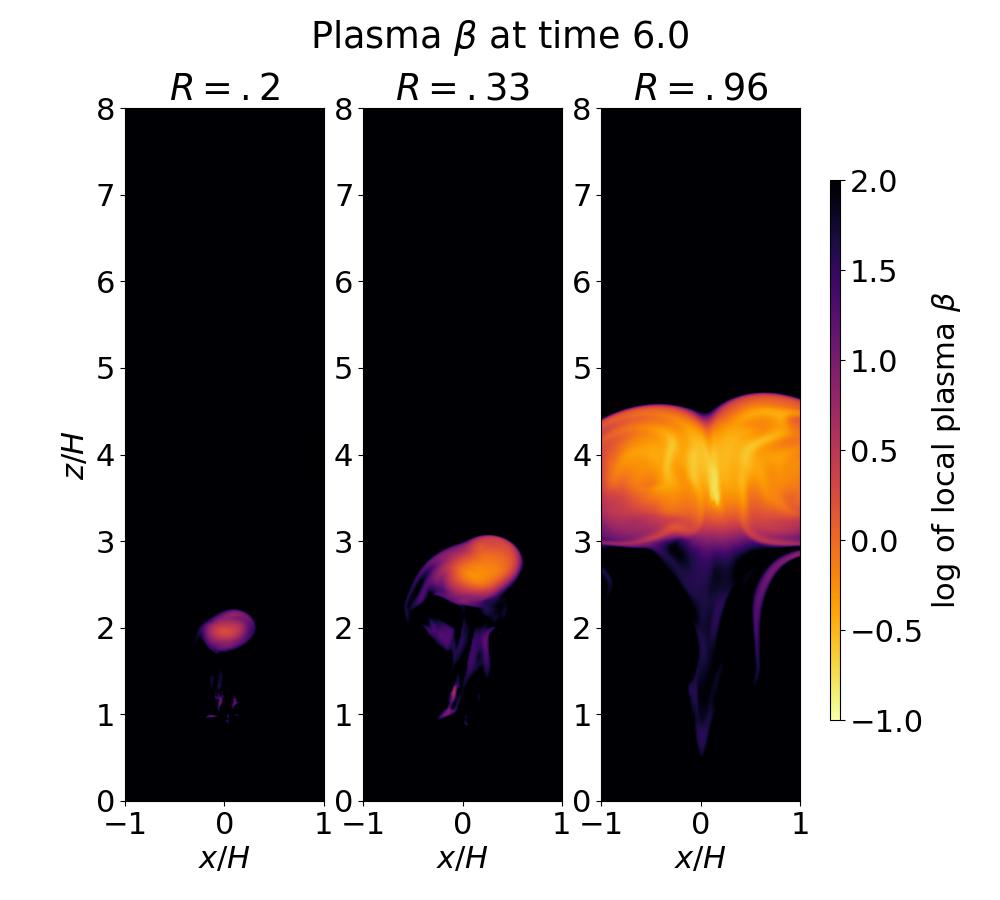}
    \end{subfigure}
    \begin{subfigure}{.49\textwidth}
        \includegraphics[width=\linewidth]{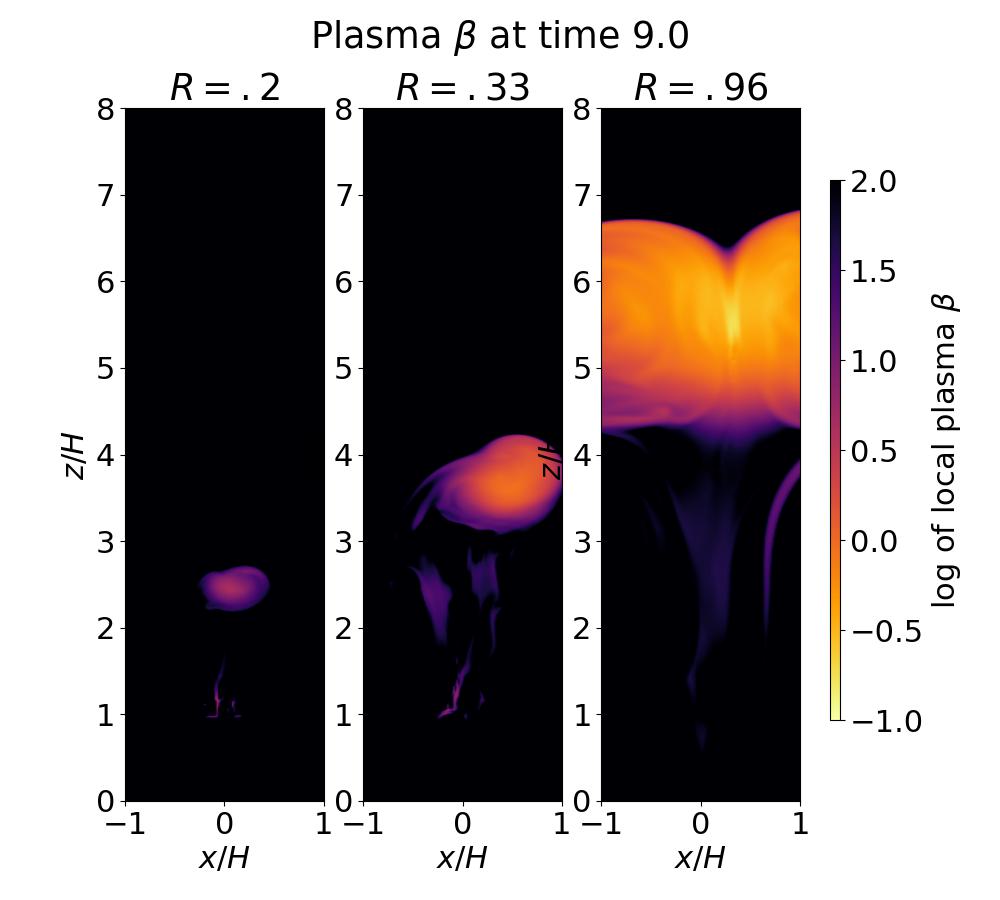}
    \end{subfigure}
    \caption{Bubble dynamics depending on the initial size, vertically aligned case, no outside \Bf. Plotted are midplane slices of $\log(\beta)$ for three cases of varying initial radius $R=0.2, \, 0.33,\, 0.96$, at times $0,3,6,9$ $\times H/c_s$.}
    \label{fig:horse_race}
\end{figure*}

We initially consider a vertically aligned bubble, Fig. \ref{fig:horse_race}, plotting $\log{\beta}$, with varying radii. The rising speed of the bubble is dependent on the initial size of the bubble, and the bubble also tilts on its side as it moves, approaching nearly horizontal.

\begin{figure*}[h!]
    \centering
    \includegraphics[width=.8\textwidth]{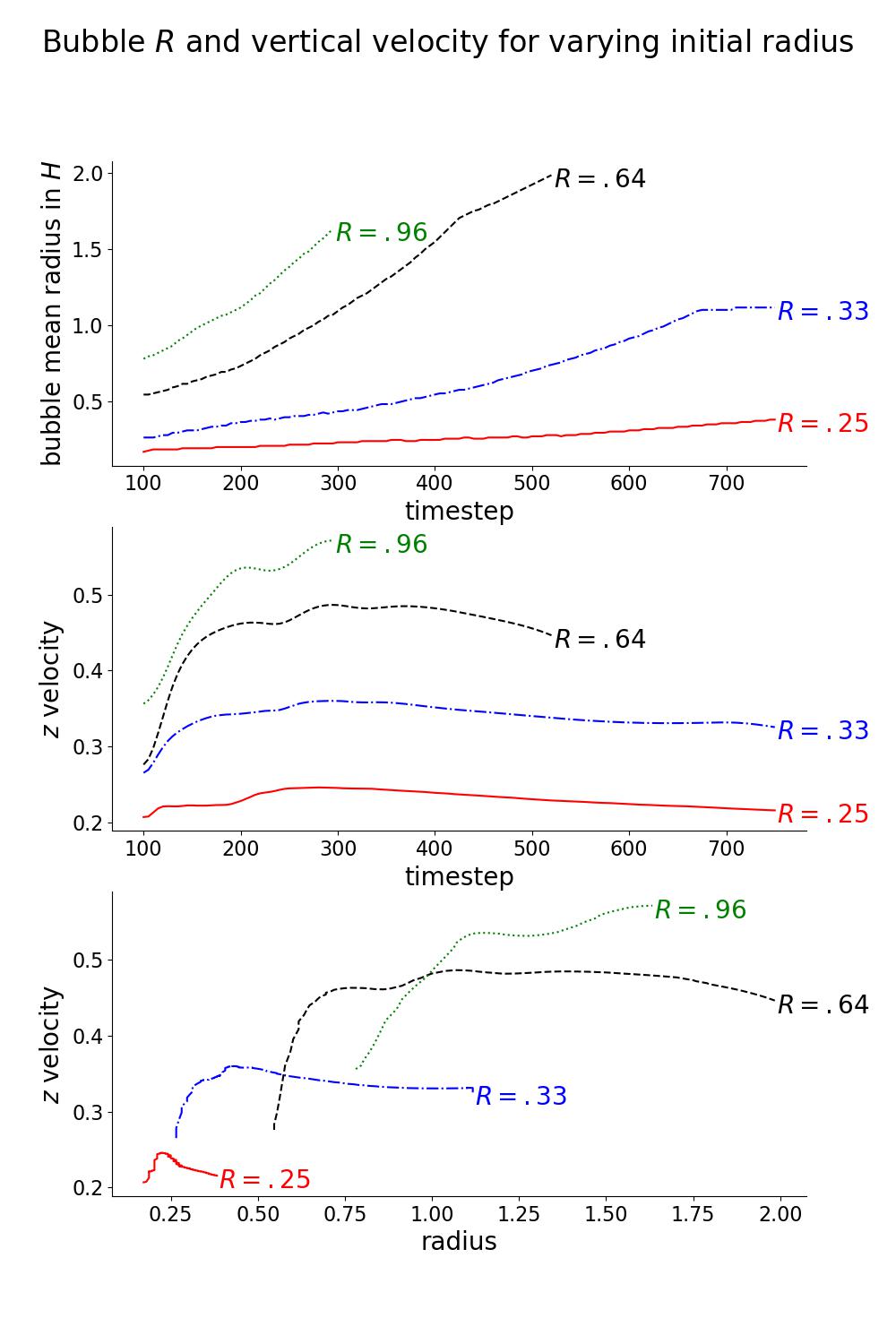}
    \caption{Location and rising speed for vertically oriented bubbles of various initial radius (scales as $(5.76/\alpha) *H $). After an initial equilibration period, the bubbles approach a constant rising velocity, with the larger bubbles rising faster. The largest bubble ($R=.96$) encounters a plunger like effect that inhibits its rising as it runs against the side walls of the domain past timestep 600. We include an intermediate size run ($R=.64$) to fill out the parameter space.}
    \label{fig:rising_speeds}
\end{figure*}

The bubbles initially accelerate from the buoyant force.  The velocity levels off, Fig \ref{fig:rising_speeds}, until the bubble impacts the reflecting boundary condition at the top, or encounters the outflow condition on the side. After $10H/c_s$ the reflections off top boundary condition and side effects become relevant, so the simulations are terminated. All of the rising speeds are much slower than the low amplitude sound waves generated by the initial equilibration of the bubble with its surrounding atmosphere. These sound waves appear to have minimal impact on the bubble dynamics on the relevant timescales. The $\alpha_{17} (R=.33)$ run drifts off to one side of the domain before interacting with the outflow boundary condition there. Notably the larger bubble expands much more relative to the others. As the plasma $\beta$ is approximately constant between runs, we see more of the fluid-dynamics dominated regime in the larger bubble, however the bubble is able to stay coherent and surface instabilities are suppressed.

As in Fig. \ref{fig:rising_speeds}, we see that, especially for the smallest bubble, the rising velocity quickly approaches a very stable equilibrium, only eventually affected by boundary conditions. We also note that the larger bubbles rise faster. Theory predicts a relationship $V/c_s\propto\sqrt{R/H}$, and an eventual constant rising velocity. We find this to be the case. The larger bubble cases continue to occasionally encounter boundary effects. It is possible that the tailing off of mean vertical velocity even in $\alpha_{17}, R=.33 H$ case is due to encountering the side boundary. As with many potential conflicts, a larger domain (and requisitely more computational time), can suppress these issues. The same is true of the $\alpha_6, R=.96$ run, where the larger bubble required a lower resolution to stay within computational limits. The rising velocity does not appear to be affected by the initial orientation of the bubble, with the $z$-oriented and $x$-oriented runs of the same $R$ rising at the same speed. 

Regardless of orientation, the bubble flattens somewhat as it rises. Similar to a falling leaf in Earth's atmosphere, we observe some side-to-side movement, resulting on some of the bubbles encountering the edges of their respective domains. The numerical characterization of this motion would require wider domains, so as to have the effect be independent of the boundary conditions chosen.

\begin{figure*}[h!]
    \centering
    \begin{subfigure}{.32\textwidth}
        \includegraphics[width=\linewidth]{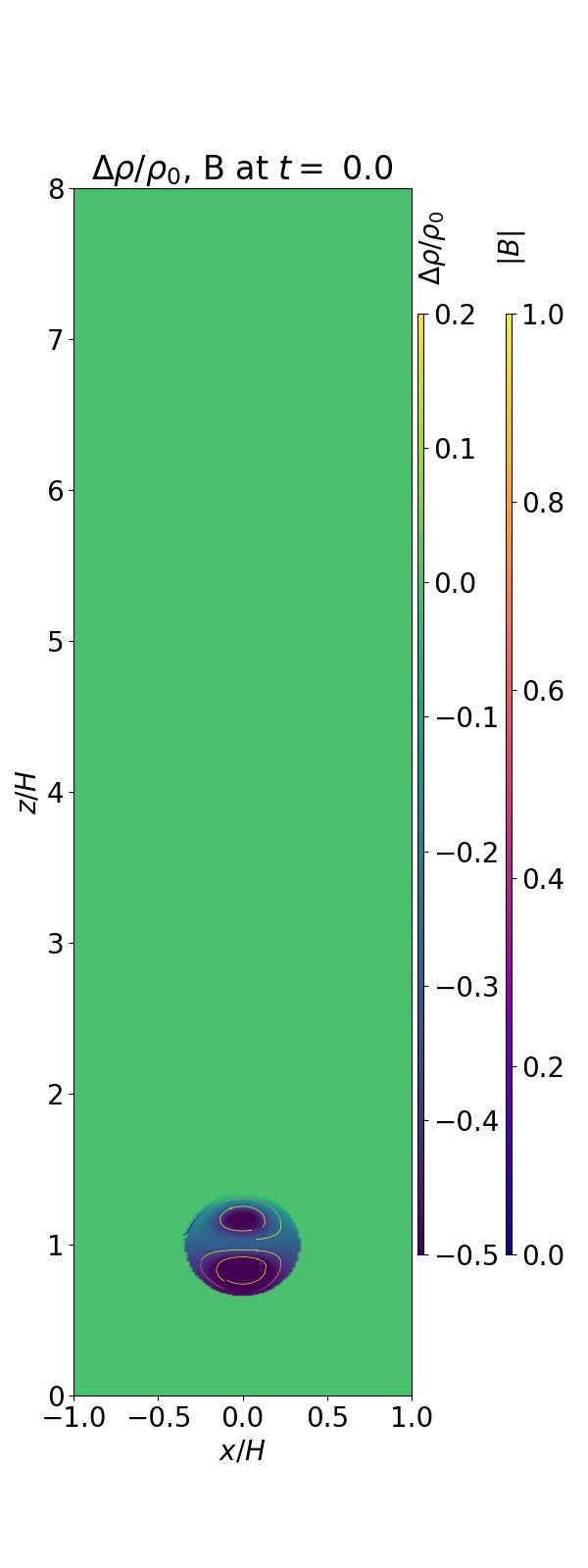}
    \end{subfigure}
    \begin{subfigure}{.32\textwidth}
        \includegraphics[width=\linewidth]{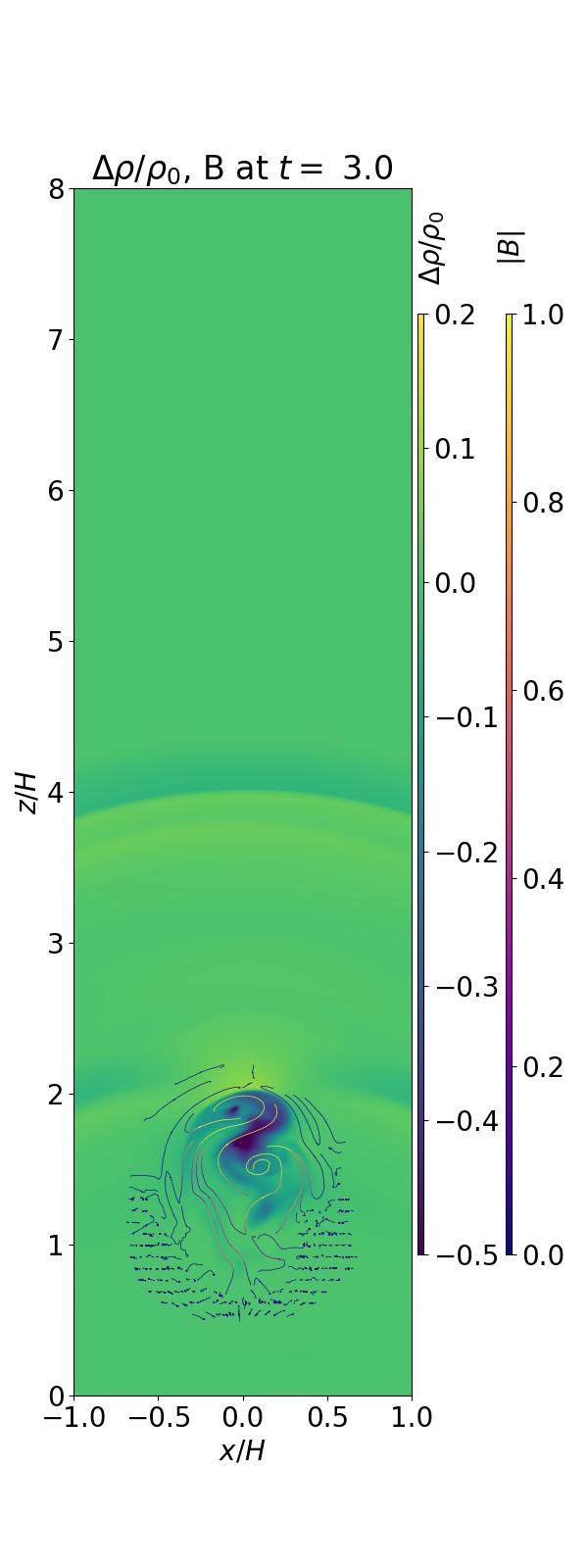}
    \end{subfigure}
    \begin{subfigure}{.32\textwidth}
        \includegraphics[width=\linewidth]{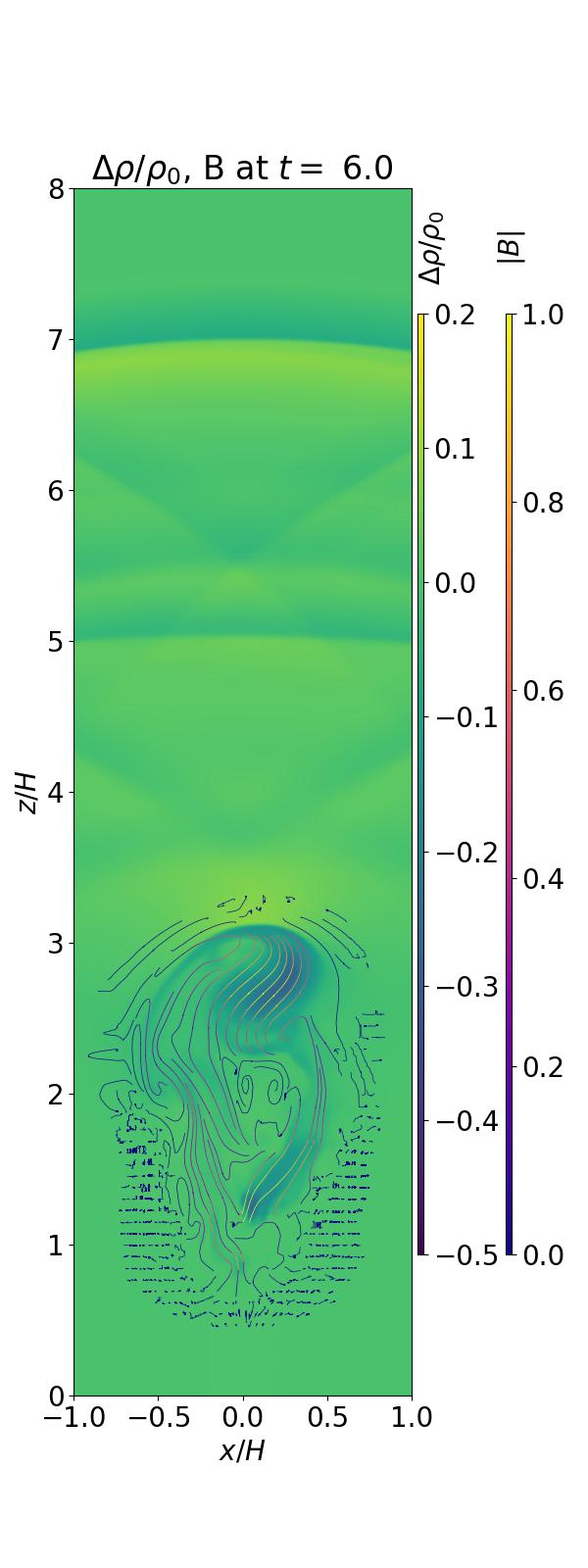}
    \end{subfigure}
    \caption{ Horizontally oriented bubble, $R=0.33$. $\Delta \rho/\rho$ and $B$ field streamlines for an initially $x$ oriented  bubble at dynamical times of 0, 2.4 and 5.1. Streamline color indicates $B$ magnitude. At these times, the rarefaction wave has yet to interact with the bubble. This is the same radius as the middle panels of \ref{fig:horse_race}. }
    \label{fig:x_orient}
\end{figure*}
Next, we simulate a horizontally aligned bubble ($x$-oriented), Fig. \ref{fig:x_orient}, with the median $R=.33$. The change in orientation does not appear to significantly affect the generation of small sound waves or the rising velocity. The critical difference between the horizontal and vertical cases comes from the relative lack of tilting. Where the $z$-oriented bubble tilts until being nearly horizontal, the initially $x$-oriented bubble remains nearly constant in its orientation until impacting the side of the domain. Unlike the $z$=oriented case, we also see compression of fluid on top of the bubble as it moves, leading to a mild overdensity there. 

We plot different quantities for the vertically and horizontally oriented bubbles, but the conclusions are the same: coherence that is sustained through motion, and suppression of surface in stabilities. The results of our work can be safely generalized to varied initial orientations of the bubble, with very similar results.

\subsection{Extension to more complicated spheromak-like shapes}

Up until now, all of our spheromaks have used the first zero of the governing magnetic field equations (See Appendix \ref{pressure-confined}). The force-free state can be extended to later zeroes of the functions as well, resulting in a "nested" spheromak shape, with one force free configuration contained inside another. 

\begin{figure*}[h!]
    \centering
    \begin{subfigure}{.3\textwidth}
        \includegraphics[width=\linewidth]{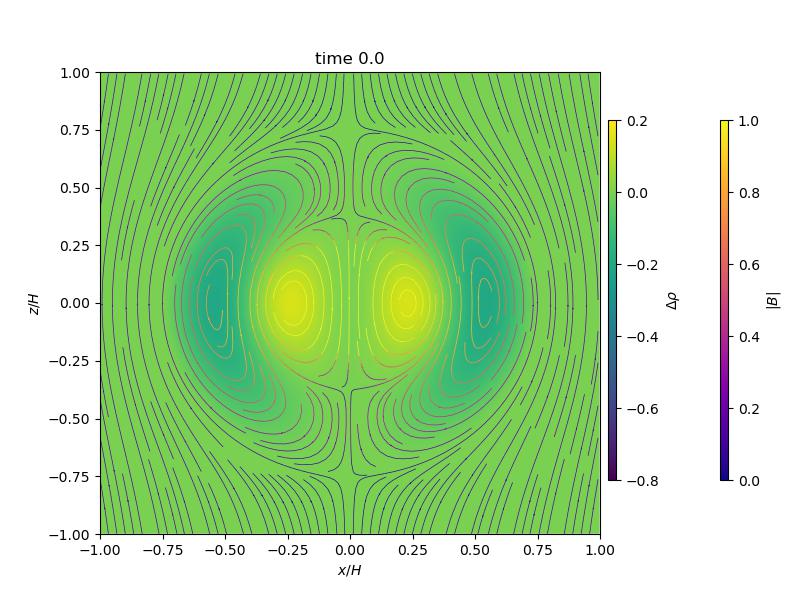}
    \end{subfigure}
    \begin{subfigure}{.3\textwidth}
        \includegraphics[width=\linewidth]{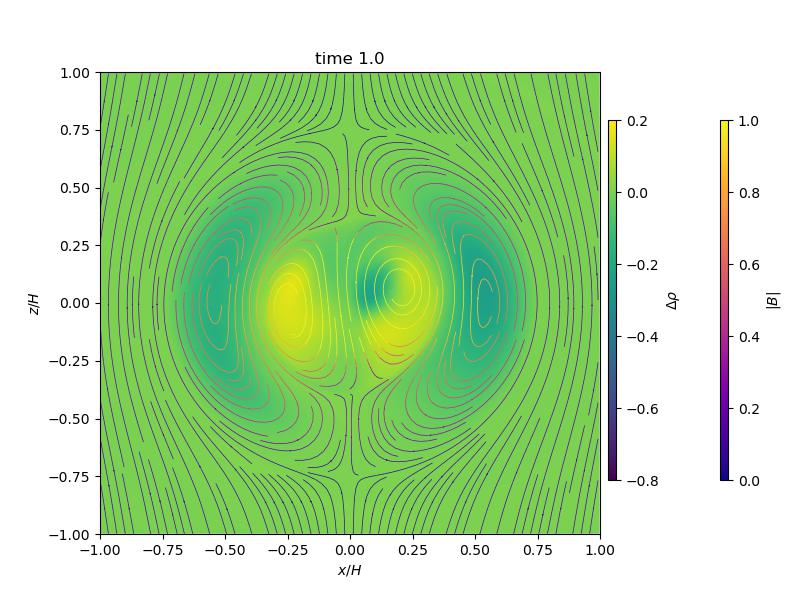}
    \end{subfigure}
    \begin{subfigure}{.3\textwidth}
        \includegraphics[width=\linewidth]{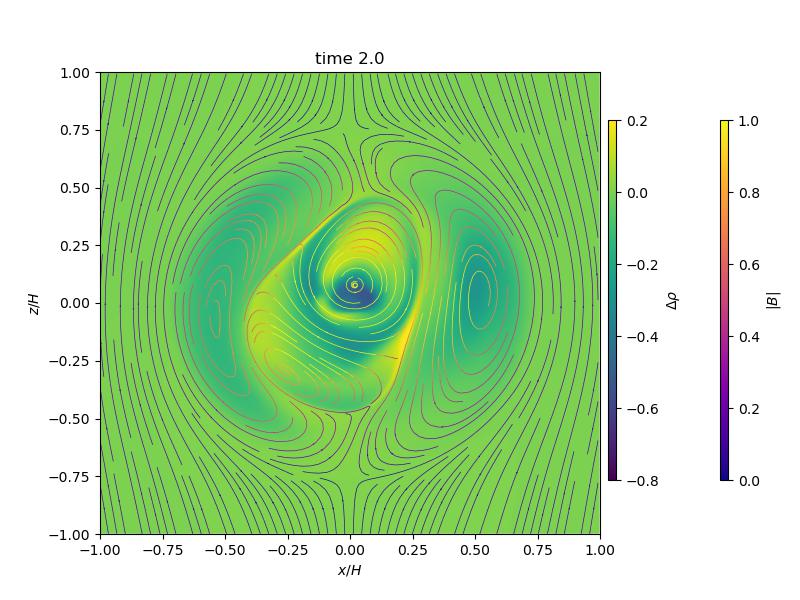}
    \end{subfigure}
    \begin{subfigure}{.3\textwidth}
        \includegraphics[width=\linewidth]{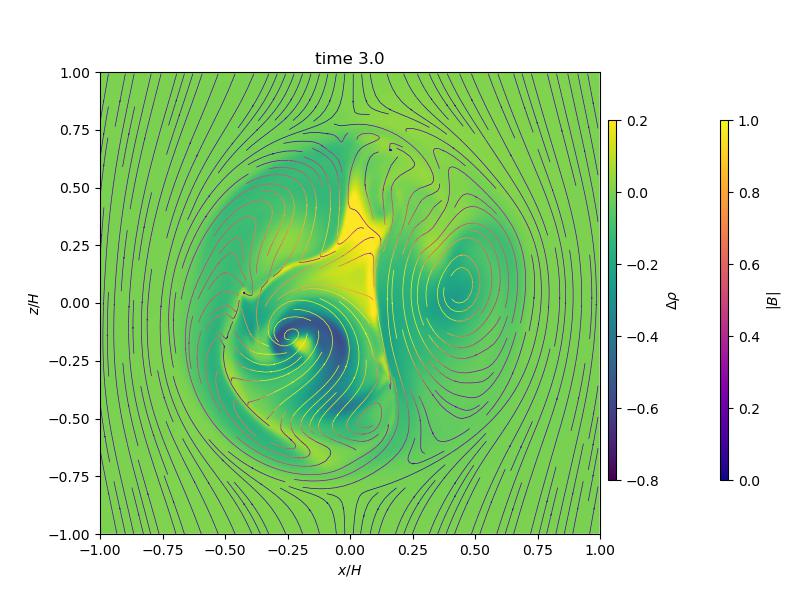}
    \end{subfigure}
    \begin{subfigure}{.3\textwidth}
        \includegraphics[width=\linewidth]{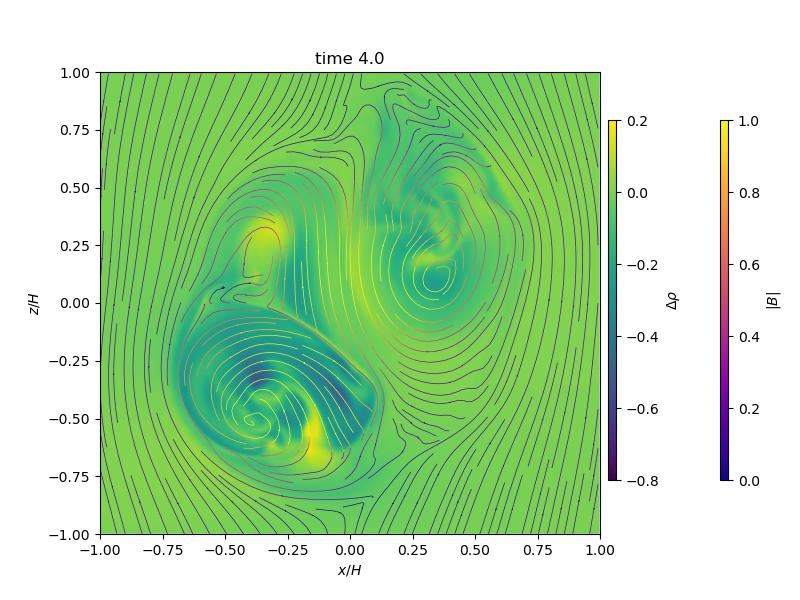}
    \end{subfigure}
    \begin{subfigure}{.3\textwidth}
        \includegraphics[width=\linewidth]{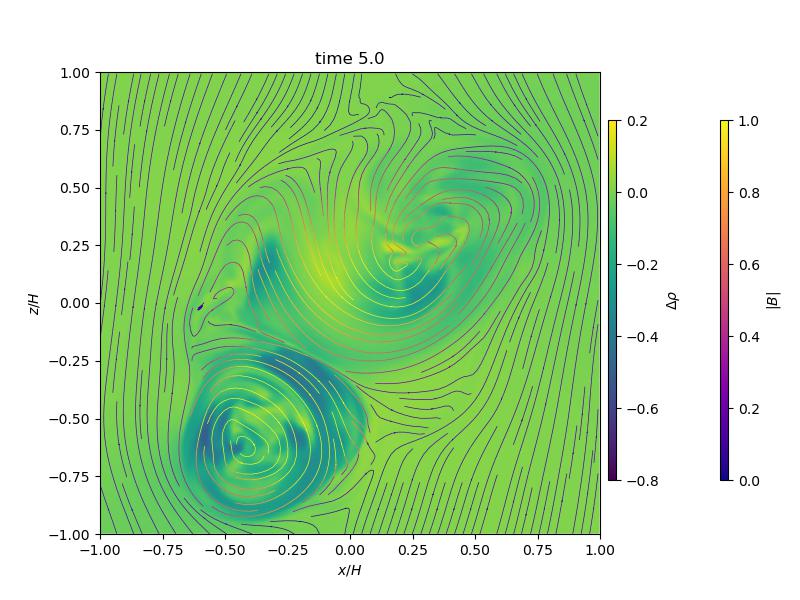}
    \end{subfigure}
    \caption{$\rho$ and $B$ field streamlines for a nested spheromak, where $R$ and $\alpha$ are decoupled from each other. After $5 H/c_s$, the 'core' spheromak has already been ejected, and a semi-stable single spheromak reached. Note there are outflow conditions on every boundary and no gravity}
    \label{fig:nested}
\end{figure*}

Maintaining the force balance inside the bubble, analogous to the $0$ surface current case, the initial condition becomes the first subplot of \ref{fig:nested}. In principle, one can layer increasingly more spheromak $B$ field shells inside the bubble, whilst maintaining force balance. However, when we include just one shell outside an initial core, the core is quickly ejected, recovering approximately a simple spheromak of the type presented earlier, in agreement with (\cite{2020JPlPh..86d9007M}). 
This run also highlights that the spheromak tilts, even if gravity is absent.

\section{Wake dissipation and the problem of reheating the cluster cores}

One of the key  problems in physics of intercluster medium (ICM)
is the absence of strong cooling flows at the centers of galaxy clusters
\citep[][]{Peterson06}. 
It has been proposed that  heating of ICM by Active Galactic Nuclei (AGNs)
may be sufficient to offset the cooling \citep{begelman04,2007ApJ...668L...1L}.
One the problems is that these  AGNs blown bubbles expand, typically, 
{\it subsonically}, as is indicated by the general absence of  shock signatures
ahead of the  bubbles.

It was suggested that dissipation in the wake of rising bubbles may contribute to reheating of ICM \cite{2018MNRAS.478.4785Z,Congyao}. Our results lend some support to this possibility. 
Consider a jet of power $L_j$, operating for time $t_j$, so that the total injected energy is $E_j = L_j t_j$. After termination and relaxation, the resulting bubble has a size given by pressure balance
\be
E_j = p_{ICM} R_0^3 \approx \rho _{ICM} c_S^2 R_0^3
\ee

As the bubble rises, the power dissipated in the wake can be estimated as 
\be
P_{dis} \approx \rho _{ICM} R_0^2 V^3
= \frac{c_s \sqrt{R_0} }{H^{3/2}} E_j
\ee
Estimating time to float a scale height as
\be
\tau_H \approx \frac{H}{V} = \frac{H^{3/2} }{c_s \sqrt{R_0}},
\ee
the dissipated energy after floating a scale height is 
\be 
E_{diss} \approx  P_{dis} \tau_H \approx E_j
\ee

We arrive at an important results: after floating one scale height, a bubble dissipates approximately its  initial  energy. Importantly, this is done gently, by a subsonically moving cavity. This Reynolds drag dissipation mechanism acts in addition to purely hydrostatic one (as a light bubble rises, heavier outside medium will fall down to fill the space, dissipating gravitational energy).

We leave a more detailed investigation of the wake dissipation to a subsequent paper. 


\section{Conclusions}

We demonstrate that low density bubbles observed in clusters of galaxies contain   internal large scale \Bf. The \Bfs\ stabilize  the bubble   against surface instabilities, that were shown to be destructive for  purely fluid bubbles. 

As a major initial assumption, we start with a relaxed internal configuration, close to the Woltjer-Taylor state.  We do not consider the process of relaxation. This can be a topic of a future numeriacl investigation. Relaxation  should occur on several (perhaps many) internal \Alfven time scales.  Woltjer-Taylor relaxation principle requires that the pre-relaxed state contains non-zero helicity. In the case of astrophysical AGN-blown bubbles, this implies that the simulation of the original jet must include the injection of both toroidal and poloidal components of the \Bf. 

Differing from a purely magnetically dominated bubble, our simulations did not show a stationary steady end state (hovering). Instead, we find a balance between the purely magnetically dominated behaviors (constant bubble size, hovering upon reaching an equilibrium point) with the purely hydrodynamic expectations (expansion and continued rising). The combined dynamics of these two paradigms and the interplay between them determines the behavior of the physical system.

We show the ability of an embedded twisted spheromak to stabilize rising plasma bubbles in a constant gravitational field. Accompanying this result, we explore the auxiliary dynamics, including rising velocities and tilting of the bubble.

Sheared/vortical velocity structure of the purely hydrodynamic  bubbles may also contribute to stability \citep{2000A&A...356..788C}. Hydrodynamical vorticity then needs to be generated through non-barotropic $\nabla p \times \nabla \rho $ term. In fact, fluid ring vortices have a lot in common with the magnetic spheromak structures (\eg, spheromak magnetic structure matches velocity structure of Hill’s vortex with a swirl \citep{Lamb}. Qualitatively, in this application, two advantages of magnetic structures versus internal velocity shearing are: (i) magnetic spheromak-like structures naturally appear as a result of Woltijer-Taylor relaxation; (ii) large scale magnetic fields also acts as a stabilizing agent against surface instabilities (through hoop stresses). We expect that magnetic fields could work in concert with other possible stabilizing mechanisms such as anisotropic viscosity \citep{2019ApJ...883L..23K} and early infaltion leading to a dense shell \cite{2006MNRAS.371.1835P}. 

While preceding observations come primarily from ChandraX, we anticipate the use of this work for other X Ray missions as well, notably JAXA's XRISM satellite. Among XRISM's stated aims is investigating the formation of Galaxy Clusters, by measuring the temperatures and velocities of the ICM. This work provides expectations and constraints on the bubble dynamics that XRISM will observe in Perseus and others. Observations indicative of large scale magnetic fields, and the fluid motions associated with them should provide ample data for evaluation of this model, and comparison with other possible stabilizing mechanisms.

We note that the stability of plasma bubbles, and spheromak-like magnetic fields is of critical importance to the development of fusion reactor tokamaks. Understanding the conditions under which such fields can be maintained and any dynamics thereof is a necessary prerequisite for containing hot plasma within, a bubble analogous to those seen in the Perseus Cluster and others. We contend the physics determined to be applicable to the latter is also of use to the former. Notably, the bubbles simulated here begin from a torus shape well known in tokamak research, and from 3-d visualizations, we know this toroid bubble can be maintained.

We foresee a number of follow-up investigations: (i) relaxation of a newly created bubble to the  Woltjer-Taylor state in a stratified external medium; (ii) effects of \Bf\ 
in the ICM: draping \citep{2006MNRAS.373...73L,2008ApJ...677..993D} and related  formation of a depletion layer,  corresponding  polarization and rotation measure maps;
(iii) interaction of a rising bubble with cluster shocks and sound waves; (iv)  more realistic equations of state,  different for external and internal  media; (v) full 3D geometry, including the density/pressure structure of a cluster; (vi) studies of wake dissipation in stratified ICM medium;
(vii) more detailed investigation of internal magnetically-driven flows;(viii) mapping of simulation results to concrete observables. Acknowledging the limits imposed by computation time, future work should employ stronger resources to address e.g. the boundary condition concerns. We also seek to expand the simulations to adiabatic atmospheres (requiring a different background polytrope), and adding tracers to the bubble. Continued investigation of tilting remains a priority, as well as a wider parameter space of starting angles,rather than only increments of $90 \degree$. The basic case of a spheromak-like large scale magnetic field stabilizing a gas bubble for many time scales opens a myriad of possibilities for exploration and explanation of physical phenomena.

The authors acknowledge insightful discussion and comments from collaborators, in no particular order: Eugene Churazov, Maxim Barkov, Sebastian Heinz, Yiting Wang, Maxim Markevitch, and Irina Zhuravleva. 
We also thank organizers of the conference McNamara@65: Understanding Feedback in Galaxies and Clusters, where preliminary results were presented.

\bibliographystyle{aasjournalv7}
\bibliography{sample,BibTex}

\appendix

\section{Initial configuration: pressure-confined magnetic bubble}
\label{pressure-confined}

As the  initial state we adopt the configuration of pressure-confined magnetic bubble \citep{2010MNRAS.409.1660G}, Fig.  \ref{fig:overview}. Below we also generalize solutions to finite surface fields.

The internal  structure of the bubble resembles spheromak configuration, but it is not force-free.   Such a setup can be understood as the result of Woltjer-Taylor Relaxation (\citep{Woltjer58,1974PhRvL..33.1139T}) in the magnetized plasma to a state satisfying the Grad-Shafranov equation: $j\times\B=\nabla p$. If there is a constant gas pressure profile, this becomes simply $\nabla \times \B = \alpha\B$, which admits the typical spheromak solutions. These spheromak-like configurations have a distinct property that total \Bf\ - both toroidal and poloidal components - vanishes on the surface. There are no current sheets. To compensate, the density must vary within the bubble such that $-\nabla P=-\nabla B^2/2+(B\cdot\nabla)B$. The requisite shape leaves a toroid depression in the fluid density which is the observable density deficit, i.e. bubble. This required balance necessarily sets a cap on how strong the magnetic field can be such that the density dip can compensate for the magnetic force. Thus the field we use is a reasonable relaxed result from magnetized astrophysical plasma, rather than an artificial construction.

 
An internal magnetic field adds a new dimension of complexity, if compared with purely hydrodynamic bubble. Anisotropic internal magnetic and gas pressure affect the overall shape of the bubble, which in turn affects its dynamics. For theoretically simple models of axisymmstric magnetized bubble, the total pressure is minimal near the axis. As a result, one expects that with time  the  shape becomes oblate. Motion of oblate objects along its axis is hydrodynamically unstable - one then expects a tilt, and resulting oscillating trajectory. This is indeed seen in our simulations. Additionally, magnetic stresses induce internal, external  flow-independent, dynamics. In Appendix  \S \ref{Suydam} we discuss weak (small growth rate) instability of a magnetically confined bubble. One of the consequences is that weak non-axisymmetric perturbations, induced by purely internal dynamics, distort the overall shape;   this is then amplified by the external drag forces, inducing   zig-zag trajectory of a bubble (a visual analogy is the trajectory of a falling leaf). We see the tilting in all cases, and for a variety of numerical setups (including orientation and resolution), regardless of whether gravity is included. It is especially visible in the smaller bubbles, e.g. $R=.33$.

Here for completeness we outline the structure of pressure-confined magnetic bubble, first derived in \cite{2010MNRAS.409.1660G}. We then argue that the structure is likely to be weakly unstable to kink/interchange instabilities.

Consider a spherical bubble of size $R$ confined by a medium with pressure $p_0$. At any point inside
\ba &&
p=p_0 +  {F_0} \Psi
\nn &&
\Psi = f(r) \sin^2 \theta
\nn &&
\B = \nabla \Psi \times \nabla \phi + \alpha \Psi \nabla \phi 
\ea
($\Psi $ is the magnetic flux function).

Force-balance requires
\be
\curl \B \times \B = \nabla p
\ee
which gives 
\ba &&
f= 
-\frac{F_0 r^2}{\alpha ^2}+c_1 R^2  j(x)  (\alpha  r)
\nn &&
 j(x) = \cos x - \frac{\sin x}{x}
   \ea
   (normalization of $F_0$ and $c_1$ are chosen in such a way that they are independent of $\alpha$).
   The  condition of zero radial field on the surface requires
   \be
   F_0 =  c_1 \frac{\alpha ^2}{R^2}  j(\alpha  R)
   \ee   
      (This also ensures that the toroidal component is zero on the surface). 
      
      Thus, generally, \Bf\ is 
      \be
  \B=    \left\{-\frac{2 \cos (\theta ) \left(r^2 j(\alpha  R)-R^2 j(\alpha  r)\right)}{r^2},\frac{\sin
   (\theta ) \left(2 r j(\alpha  R)-\alpha  R^2 j'(\alpha  r)\right)}{r},-\frac{\alpha  \sin
   (\theta ) \left(r^2 j(\alpha  R)-R^2 j(\alpha  r)\right)}{r}\right\}
   c_1
   \label{Bbubble}
   \ee
Conditions $B_r(R)=0$ and $B_\phi(R)=0$ are satisfied by construction.
Since 
\be
B_\theta (R) =  c_1 \left( 2 j(\alpha  R)-\alpha  R j'(\alpha  R)) \right) \sin \theta/R
\ee
we can set the surface field to zero by choosing 
\ba && 
2 j(\alpha  R)-\alpha  R j'(\alpha  R)) =0  \to    \tan (\alpha  R)=\frac{3 \alpha  R}{3-\alpha ^2 R^2}
\nn && 
  \alpha  R=5.763
  \ea
  gives zero field on the surface.    In this case, all the fields are zero on the surface.

  In fact, 
   we can embed the magnetic bubble in any arbitrary external \Bf, positive or negative. Let's parametrize that external field by its value on the surface $B_0$, so that
   \ba &&
   \B_{\rm in} =  \left\{2 \cos (\theta ) \left(j(\alpha  R)-\frac{R^2 j(\alpha  r)}{r^2}\right),\frac{\sin (\theta
   ) \left(\alpha  R^2 j'(\alpha  r)-2 r j(\alpha  R)\right)}{r},\frac{\alpha  \sin (\theta )
   \left(r^2 j(\alpha  R)-R^2 j(\alpha  r)\right)}{r}\right\}  
   \times
   \nn &&
    \frac{B_0} { \alpha  R j'(\alpha  R)-2 j(\alpha  R)}  
    \ea

   The resulting structure is spheromak-like, but is  not force-free.
   Kinetic pressure is 
   \be
   p=p_0 + \frac{\alpha ^2 c_1{}^2 \sin ^2(\theta ) j(\alpha  R) \left(R^2 j(\alpha  r)-r^2 j(\alpha 
   R)\right)}{R^4} = \frac{\alpha ^2 B_0^2 \sin ^2(\theta ) j(\alpha  R) \left(R^2 j(\alpha  r)-r^2 j(\alpha 
   R)\right)}{R^4 \left(\alpha  R j'(\alpha  R)-2 j(\alpha  R)\right)^2}
   \ee
   (second form assumes $B_0 \neq 0$, non-zero denominator.)
      It reaches minimum at 
   \be
   j'(\alpha  r)=-\frac{2 r j(\alpha  R)}{\alpha  R^2}
\ee
For special case of  zero surface field,  $r_{min} = 0.5130 R. $

\section{Weak internal  kinking instability of pressure-confined magnetic  bubble}
\label{Suydam}
The basic pressure-confined spheromak discussed in \S \ref{pressure-confined} is  weakly unstable. This follows from Suydam's criterion \cite{Suydam,1968JETP...26..682S,1973PhFl...16.1927G}  that describes  stability against  local interchange instabilities, originally  in  cylindrical pinch. As an estimate/approximation we can the relations in the equatorial plane of the bubble, especially close to the axis, where approximation of cylindrical pinch is likely to be better satisfied (indeed, as we show below, the instability occurs close to the axis). 

\begin{figure}[h!]
\includegraphics[width=.99\linewidth]{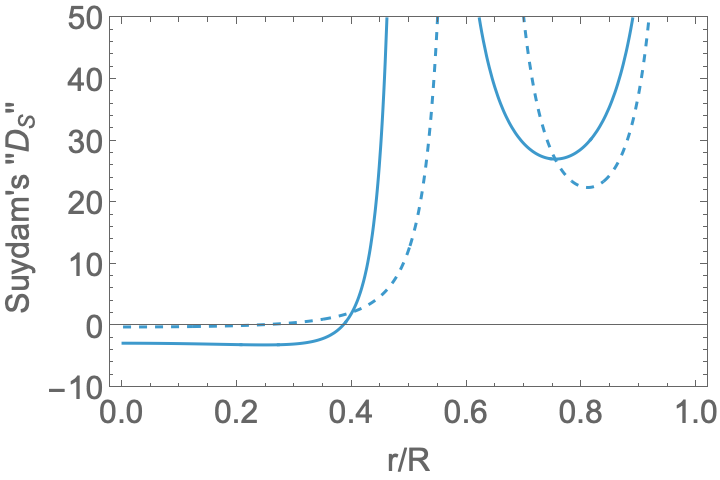}
\caption{Suydam's $D_S$ evaluated in the equatorial plane of pressure-confined spheromak (solid line). Negative values imply instability to  local  kink/interchange perturbations near the axis.
Dashed line: force-free spheromak -  in this case $D_S\geq 0$ everywhere.}
\label{SuydamS}
\end{figure}

Suydam's  stability criterion reads
\ba&&
D_S=\frac{1}{4} \left(\partial_\varpi \ln q(\varpi ) \right) ^2+
\frac{ 2 \partial_\varpi  p}{\varpi  B_z^2} > 0
\nn &&
q=  \varpi B_z/B_\phi
\ea
($\varpi $ is cylindrical radius). The form of $D_S$ indicates that the criterion describes pressure-driven (not current-driven) instabilities, typically on the slow MHD modes. Given the magnetic structure of a bubble (\ref{Bbubble}), we can evaluate Suydam's  $D_S$ in the equatorial plane, Fig. \ref{SuydamS}. Near the axis Suydam's  $D_S$ is negative indicating  weak instability. (In dimensionless units the asymptotic values $D_S(r\to 0) = -2.63$, much smaller that  typical positive value in the bulk of $\sim$ dozens.) For force-free spheromak Suydam's  $S$ is naturally positive everywhere, except on the axis where it is zero. 

Our numerical results are consistent with this conclusion. Our very long run of isolated pressure-confined spheromak (without external pressure gradient) show what  seems to be slowly developing instability, Fig. \ref{fig:meshlab} Due to slow growth rate the instability was missed by \cite{2010MNRAS.409.1660G}. Our results are consistent with the general concept that violating Suydam criterion produces feeble instabilities \citep{2002PhPl....9.3395G}.  After sufficiently long times the core falls to kink instabilities.
The growth rate of the instability in our case
is far slower than any of the processes we are interested in, so it's not likely to affect our conclusion.

\end{document}